\documentclass[twocolumn,showpacs]{revtex4}

\usepackage{graphicx}%
\usepackage{dcolumn}
\usepackage{amsmath}
\newcommand{\be}{\begin{equation}}
\newcommand{\ee}{\end{equation}}
\newcommand{\bey}{\begin{eqnarray}}
\newcommand{\eey}{\end{eqnarray}}
\newcommand{\bw}{\begin{widetext}}
\newcommand{\ew}{\end{widetext}}
\newcommand{\bi}{\begin{itemize}}
\newcommand{\ei}{\end{itemize}}
\newcommand{\bem}{\begin{enumerate}}
\newcommand{\eem}{\end{enumerate}}
\newcommand{\ww}{\widetilde}

\newcommand{\ra}{\rangle}
\newcommand{\la}{\langle}
\newcommand{\pp}{\partial}
\newcommand{\br}{ {\bf r} }
\newcommand{\bp}{ {\bf p} }

\begin{document}

 \title {
 Stability of quantum motion in regular systems: a uniform
 semiclassical approach
 }

 \author{Wen-ge Wang$^{1,2}$, G.~Casati$^{3,4,1}$, and Baowen Li$^{1,5,6}$}

 \affiliation{
 $^1$Department of Physics and Centre for Computational Science and Engineering,
 National University of Singapore, 117542, Republic of Singapore
 \\ $^{2}$Department of Physics, Southeast University, Nanjing 210096, China
 \\ $^{3}$Center for Nonlinear and Complex Systems, Universit\`{a}
  degli Studi dell'Insubria, Via Valleggio 11, 22100 Como, Italy
 \\ $^4$CNR-INFM and Istituto Nazionale di Fisica Nucleare, Sezione di Milano, Italy
 \\ $^5$Laboratory of Modern Acoustics and Institute of Acoustics, Nanjing University, 210093, China
 \\  $^6$NUS Graduate School for Integrative Sciences and Engineering, National University of Singapore,
 117597, Republic of Singapore
 }

 \date{\today}

\begin{abstract}

We study the stability of quantum motion of classically regular systems in presence of small perturbations. On
the base of a uniform semiclassical theory we derive the fidelity decay which displays a quite complex
behaviour, from Gaussian to power law decay $t^{-\alpha }$  with $1 \le \alpha \le 2$.
 Semiclassical estimates are given for the time scales separating the
 different decaying regions and numerical results are presented which confirm our theoretical predictions.

\end{abstract}

\pacs{05.45.Mt, 03.65.Sq }

\maketitle



 Stable manipulation of quantum states is of importance in many research fields
 such as in quantum information processing
 and in Bose-Einstein condensation.
 A measure of the stability of quantum motion is
 the so-called fidelity or quantum Loschmidt echo \cite{Peres84}, which characterizes the stability of quantum dynamics
 under small perturbations of the Hamiltonian that may derive from static imperfections or from
 interaction with an external environment. From a more general point of view it is interesting to
 understand the behavior of fidelity in relation to the dynamical properties of the system.

 While fidelity decay in classically chaotic systems has been extensively studied
 \cite{JP01,JSB01,CT02,PZ02,BC02,VH03,WCL04,Vanicek04,WL05,GPSZ06},
 the situation in regular systems is much less
 clear\cite{PZ02,JAB03,PZ03,SL03,Vanicek04,WH05,Comb05,HBSSR05,pra05-bec,GPSZ06,WB06} and
 only for the particular case of vanishing time-average
 perturbations
 a clear understanding has been achieved  \cite{PZ03}.
 Moreover, for a general perturbation and for a single initial Gaussian wave
 packet, a Gaussian decay has been predicted
 in some time interval which is not exactly specified\cite{PZ02}.
 On the other hand, numerical investigations show a much more rich behavior of the fidelity
 decay ranging from
 power law to exponential, up to a Gaussian decay, depending on
 initial conditions, perturbation strength, and time interval\cite{WH05}.
 In addition, a somehow unexpected regime in which the fidelity decay
 in regular systems is faster than in classically chaotic systems has
 been found in~\cite{PZ02}. All the above calls for a theory which
 can account for these diverse analytical and numerical findings.

 We would like to draw the reader's attention to the
 fact that in the general theory of dynamical systems, integrability
 is the exception rather than the rule and it is therefore extremely
 rare. However, the theory we present in this paper applies also to
 integrable regimes of systems with divided phase space in which both
 chaotic and integrable components are present and this is indeed the
 typical situation. Moreover as shown in \cite{bcs}, for a proper
 operability of a quantum computer it is desirable to remain below
 the border for transition to quantum chaos, a situation which is
 likely to correspond to quasi integrable behavior. Finally,
 some quantum algorithm can have a phase
 space representation and, in particular, the Grover algorithm can be interpreted as a simple quantum map
 which turns out to be a regular map \cite{saraceno}.

 The above considerations motivate our interest in the stability of integrable motion.
 In this letter, we develop a uniform semiclassical approach to the fidelity decay in regular systems
 and we provide a unified description together with the corresponding time
 scales. Numerical computations confirm our analytical estimates.

 Quantitatively, the fidelity for an initial state $|\Psi_0\ra $ is
 defined as $ M(t) = |m(t) |^2 $, where
 \be m(t) = \la \Psi_0|{\rm exp}(i Ht/ \hbar ) {\rm exp}(-i H_0t / \hbar) |\Psi_0 \ra .
 \label{mat} \ee
 Here $H_0$ and $H=H_0 + \epsilon V$ are the unperturbed and perturbed Hamiltonians,
 with $\epsilon$ a small quantity and $V$
 a generic perturbing potential.

 Consider an initial Gaussian wave packet in a $2d$-dimensional phase space,
 centered at ($\ww \br _0,\ww {\bf p}_0$),
 \be \label{Gauss-wp} \psi_0 (\br_0 ) = \left (  {\pi \xi ^2} \right )^{-d/4}
 {\rm exp} \left [ i \ww \bp_0 \cdot \br_0 / {\hbar}
 - (\br_0 - \ww \br_0 )^2/ (2 \xi ^2) \right ].  \ee
 For a sufficiently narrow, initial Gaussian packet,
 the semiclassical approximation to the fidelity amplitude is
 \bey m_{\rm sc}(t)  \simeq \left (\pi w_p^2  \right )^{-d/2}
 \int  d{\bp_0}  \exp { \left [ \frac{i}{\hbar} \Delta S
 - \frac{(\bp_0 - \ww \bp_0 )^2}{w_p^2} \right ] },
 \label{mt-sc} \eey
 where $w_p = \hbar / \xi$ and $\Delta S$ is the action difference
 between the two nearby trajectories  of the two systems $H$ and $H_0$ starting at $(\bp_0 , \ww \br_0 )$\cite{VH03}.
 We mention that, for not very narrow initial Gaussian packets, Eq.~(\ref{mt-sc}) may still hold
 with a redefinition of $w_p$\cite{WL05}.
 For more general initial states, the fidelity amplitude can be expressed in terms of the Wigner function
 of the initial state in a uniform semiclassical approach \cite{Vanicek04}.

   The action difference can be calculated in the first order classical perturbation theory:
 \be \Delta S \simeq \epsilon  \int_0^t dt' V[{\bf r}(t'),\bp (t')] \label{DS} \ee
 with $V$ evaluated along one of the two trajectories.
  Equations (\ref{mt-sc}) and (\ref{DS}) give quite accurate predictions even for relatively
 long times, much more accurate than that usually expected for a first order
 perturbation treatment \cite{VH03,WCL04}.
 The reason for the unexpected accuracy is explained in \cite{Vanicek04} by making use of the shadowing theorem,
 the two trajectories for $\Delta S$ being in fact the so-called shadowing trajectories
 of the two systems with slightly different initial conditions.

 Equation (\ref{mt-sc}) shows that the behavior of fidelity is mainly determined by
 two factors: (i) $\Delta S$ as a function of $\bp_0$ and $t$; (ii) the Gaussian term,
 which specifies an effective window in the $\bp_0$ space with size $w_p$.
 For simplicity, in what follows, we consider kicked systems with $d=1$
 and set the domains of $r$ and $p$ to be $[0,2\pi)$.

\begin{figure}
\includegraphics[width=\columnwidth]{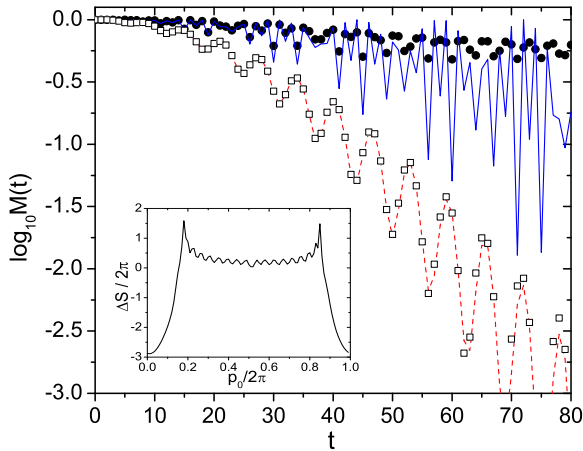}
 \vspace{-0.2cm}
 \caption{
 Comparison between the numerically computed fidelity decay ( black circles and open squares) and the semiclassical prediction
 $M_{\rm sc1}(t)$ in Eq.~(\ref{Mt-t1}), in the kicked rotator model
 with $k=0.3, \sigma =1.5, N=2^{17}$, and $\xi^2 =\hbar /20$.
 Centers of the initial Gaussian packets are: $(\ww r_0,\ww p_0)=(1.2\pi ,0.6\pi )$ for the circles,
 and $(1.2\pi ,0.2\pi )$ for the squares.
 The solid and dashed curves represent the semiclassical predictions, respectively,
 with $k_p$ evaluated numerically in the classical systems.
 Inset: $\Delta S/2\pi $ vs $p_0/2\pi$, for $r_0=1.2 \pi $ at $t=20$.
 The average slope of $\Delta S$ in the monotonically increasing part
 is much larger than that in the central part, therefore, the fidelity
 decay at $\ww p_0 =0.2\pi $ is faster than that at $\ww p_0 =0.6\pi $.
 The oscillations of $\Delta S$ in the central region imply
 larger second and higher order terms in the expansion of $\Delta S$ around $\ww p_0=0.6\pi $,
 hence, larger deviations of the fidelity from $M_{\rm sc1}(t)$.
 }
 \label{fig-mt-k03-sc}
 \end{figure}

 For an integrable system or a regular region of a mixed system, at least locally there exist
 action-angle variables $(\theta ,I)$, connected to the variables $(r,p)$
 by a canonical transformation.
 The integrand in Eq.~(\ref{DS}) can then be written as
 \be V_t \equiv  V[I(t),\theta (t)], \ \ \text{with} \ \ \ \theta (t) =\theta_0+\nu t, \label{V-t} \ee
 $I(t)\simeq I_0$,  and $\nu \equiv \pp H(I) / \pp I$.

 The main features of $\Delta S$ as a function of $t$ and $p_0$
 can be seen by substituting Eq.~(\ref{V-t}) in Eq.~(\ref{DS}),
 replacing $\nu t$ by $\phi = \nu t$, and noting the periodicity of the angle $\theta $.
 This gives
 \bey \label{DS-U} \Delta S  \simeq \epsilon (U_I t + S_f)  , \ \   \text{where}
 \  U_I \equiv \frac{1}{2\pi} \int_{0}^{2\pi} V(I,\theta )d\theta ,
 \\ \label{Sf} S_f \equiv \frac{1}{\nu } \left [
 \int_0^{b} V(I,\theta_0 + \phi ) d\phi - bU_I \right ] . \eey
 Here $b \equiv \nu t - 2\pi n_t$, where $n_t$ is the integer part of $\nu t / 2\pi $.
 Therefore, for a fixed $p_0$, $b$ is a sawtooth-type function of time oscillating between 0 and $2\pi $
 with a frequency $\nu / 2 \pi $.  As a result $S_f$ oscillates
 correspondingly,
 hence $\Delta S$ oscillates around its linearly increasing part $\epsilon U_I t$.
 On the other hand,  for a fixed $t$, $U_I$ as a function of $p_0$ changes almost
 linearly in the neighborhood of $\ww p_0$ while $S_f$
 oscillates with a frequency
 $\ww {\nu} 't /2\pi $, where $\nu ' \equiv \pp \nu / \pp p_0 $.
 Therefore the average slope of $\Delta S$ with respect to $p_0$ is $\epsilon U_I' t$,
 where $ {U_I}' \equiv {\pp U_{I}}/{\pp p_0} $.
 (We use a tilde above a quantity to indicate its value taken at
 the center ($\ww r _0,\ww { p}_0$).)

\begin{figure}
 \includegraphics[width=\columnwidth]{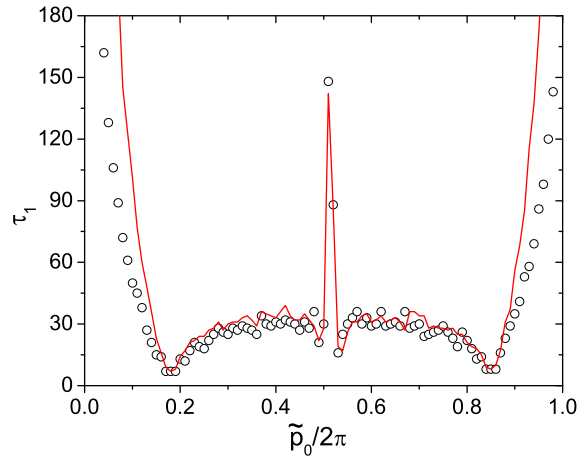}
 \vspace{0.2cm}
 \caption{
 The time scale $\tau_1$ versus $\ww p_0$, for the same parameters as in Fig.~\ref{fig-mt-k03-sc}.
 The circles give the values of $\tau_1$ calculated by the first time at which
 $|[M(t)-M_{\rm sc1}(t)]/M(t)|>0.1$ [see Eq.~(\ref{Mt-t1})].
 The solid curve is the semiclassical estimate given by
 $\sigma w_p^2 |\ww k_{pp}(\tau_1)| = 1$,  with $\ww k_{pp}(t)$ numerically computed in
 the corresponding classical system.
 } \label{fig-tau1-p03}
 \end{figure}

 We first study the fidelity decay for times $t<\tau_1$,
 where $\tau_1$ is the time scale before which the right hand side of
 Eq.~(\ref{mt-sc}) can be calculated by
 a linear approximation to $\Delta S$ with respect to $p_0-\ww p_0$.
 This gives
 \be M_{\rm sc1}(t) \simeq \exp \left [  - \frac 12 (\sigma w_p \ww k_p)^2  \right ] ,
 \ \ \text{with} \ \ k_p \equiv  \frac{1}{\epsilon } \frac{\pp \Delta S }{\pp p_0}.  \label{Mt-t1} \ee
where $\sigma = \epsilon / \hbar $ is the strength of perturbation.
 The explicit dependence of $k_p$ on time $t$ can be calculated by using Eq.~(\ref{DS-U}).
 The leading terms give
 \bey \label{kp} k_p \simeq (U_I'   + U_{\theta} )t, \  \ {\rm where} \
 U_{\theta } \equiv \frac 1t \frac{\pp S_f}{\pp p_0}  \simeq (V_t - U_I)\frac{\nu ' }{\nu }.  \eey
 Since $U_{\theta}$ oscillates in time around zero,
 the fidelity has, on average, an initial Gaussian decay
 with a rate depending on  initial conditions,
 with larger values of $|U_I'|$ implying faster decay.

\begin{figure}
 \includegraphics[width=\columnwidth]{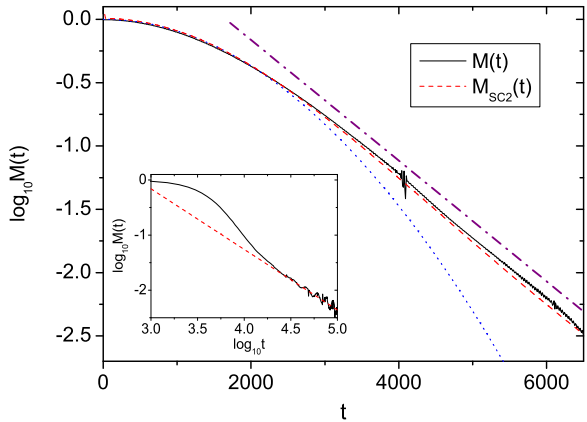}
 \vspace{0.2cm}
 \caption{
 Solid curve: Fidelity decay for  $\sigma =0.2$,
 $N=2^{12}$, $k=0.3$, $\xi^2 =\hbar /20$ and $(\ww r_0,\ww p_0)=(1.2\pi ,0.8\pi )$.
 Dashed curve: the prediction of Eq.~(\ref{Mt-sc-2}) with $c = 1$
 and $\ww U_I' \approx -0.019$, $\ww U_I'' \approx 0.042$
 numerically computed from the {\it classical} system.
 Dotted curve: the Gaussian decay $e^{-(w_p \sigma \ww U_I' t)^2/2}$.
 (For small values of $\sigma $, the difference between this decay
 and $M_{\rm sc1}$ in Eq.~(\ref{Mt-t1}) is small before $\tau_1$.)
 The dashed-dotted line shows the approximate, intermediate, exponential
 decay.
 Inset: Long time decay for the same case with $\ww r_0 = 0.6 \pi $
 (every 500 steps shown). The dashed line gives the $1/t^{1.1}$ decay.
 } \label{fig-gauss-exp-real}
 \end{figure}

 To test the above predictions, we consider the kicked rotator model,
 $  H= \frac 12 {p^2} + k_0 \cos r \sum_{n=0}^{\infty } \delta(t - n T )  \label{H} $.
 The quantized system has a finite Hilbert space with dimension $N$.
 We take $k_0 = kN/2\pi $ and $T=2\pi /N$, with $k=k_0T$ independent on $N$.
 The classical limit corresponds to $N \to \infty $.
 The one period quantum evolution is given by the Floquet operator
 $ \label{U} U = \exp [-i {\hat p}^2 T /2] \exp [-i k_0 \cos ( {\hat r}) ] $,
 and is numerically computed by the method of fast Fourier transform.
 Here $2\pi /N$ serves as an effective Planck constant.
 The inset of Fig.~\ref{fig-mt-k03-sc} shows an example of $\Delta S$ vs $p_0$,
 in which the values of $|U_I'|$ are large at the borders
 while are quite small in the central, oscillating region of $p_0$.
 In agreement with our theory, under the perturbation  $k \to k+\epsilon $,
 the fidelity for initial states lying in the two regions of $p_0$
 has a quite different decaying rate, as seen in Fig.~\ref{fig-mt-k03-sc}.
 The agreement with the theory is particularly good in the case where the
 oscillations of $\Delta S$ are not too strong.

 The time scale $\tau_1$ can be estimated by the time at which the
 second-order term in the Taylor expansion of $\Delta S/ \hbar$ is
 order of one at the point $p_0 = \ww p_0 +w_p$ , i.e.
 $\sigma w_p^2 |\ww k_{pp}(t)| \sim 1$  for $t=\tau_1$, where
 \be k_{pp}(t) \equiv \frac{ \partial k_p }{ \partial p_0} = \frac{(\nu ')^2}{\nu }
 \frac{\pp V}{\pp \theta} t^2 + O(t). \label{tau1}
 \ee
 A numerical check of this prediction is presented in Fig.~\ref{fig-tau1-p03}.
 When the $t^2$ term dominates, $\tau_1 \propto \sqrt{ 1/ \sigma w_p^2}$.

\begin{figure}
 \includegraphics[width=\columnwidth]{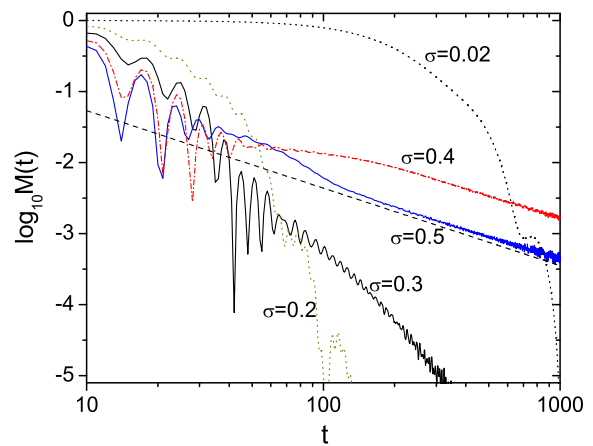}
 \vspace{-0.2cm}
 \caption{
 Fidelity decay for several values of $\sigma $ with $k=0.3$,  $N=2^{12}$, $\xi^2 =\hbar /20$,
 and $(\ww r_0,\ww p_0)=(1.2\pi , 0.28\pi )$.
 For $ t $ around 100, the rate of fidelity decay increases for $0 < \sigma < 0.2$,
 then decreases for $0.2 < \sigma < 0.4$, and then increases again.
 This is in agreement with the semiclassical prediction that the decay rate $\beta $
 oscillates with a period $|\ww \nu ' / \ww U_I'| \approx 0.4$ (see text).
 The values $\ww U_I' \approx 1.85$ and $\ww \nu ' \approx 0.76$ are obtained from the classical motion.
 Notice that the transition time from Gaussian to power law decay of
 fidelity
 is proportional to $1/\sigma $ (see text). Here the power law decay
 is visible for $\sigma =0.4$ and 0.5
 (the dashed straight line gives the $1/t^{1.09}$ decay).
 }
 \label{fig-many-ft}
 \end{figure}

 Beyond $\tau_1$, a direct analytical computation of the fidelity is
 difficult since
 higher and higher order terms in the Taylor expansion
 of $\Delta S$ with respect to $p_0-\ww p_0$ need to be considered.
 We take therefore the following approach: we divide the domain of $p_0$
 into segments separated by points $p_{0j}$ (with $p_{00}=\ww p_0$),
 in such a way that $S_f$ completes one oscillation  period within each segment.
 (We will use the subscript $j$ to indicate quantities taken at the point $p_{0j}$).
 It is easy to see that $ (p_{0j+1} - p_{0j}) \nu_j ' t \simeq 2\pi$
 and therefore the number of segments increases linearly with time
 $t$. 
 The quantity $m_{sc}(t)$ in Eq.~(\ref{mt-sc}) can now be written
 as a sum of contributions $m_j(t)$ of different segments, namely, $m_{sc}(t) = \sum_{j} m_j(t)$.

 Let us introduce the time scale $\tau_s$, at which $S_f$ completes one oscillation within
 the window   $W_p = (\ww p_0 -w_p,\ww
 p_0 +w_p)$. From Eq.~(\ref{Sf}), $ \tau_s \approx \pi / (\ww \nu 'w_p)  $.
 For $t \gg \tau_s$, there are many segments within $W_p$
 hence, within each segment, the
 variation of the Gaussian term on the right hand side of Eq.~(\ref{mt-sc})
 is negligible. Therefore one may write
 \bey \label{mj} m_j(t) \simeq \frac{e^{- (p_{0j}-\ww p_0)^2/w_p^2 }}
 {\sqrt{\pi} w_p \nu '_{j} t} e^{ i \Delta S_j /\hbar } F(t),
 \\ F(t) = \int_{0}^{2\pi } d\phi  \exp \left \{ \frac{i \sigma}{\ww \nu }
 \int_{\ww \theta (t)}^{\ww \theta (t) +\phi } V(\ww I, \theta ') d\theta '
 \right \}, \label{Ft}  \eey
 Indeed for $p_{0j}$ not far from $\ww p_0$,  $F(t)$ is independent on $j$
 and since $|F(t)|$ does not decay with time we will not consider it further.

 First we notice that equation (\ref{mj}) predicts a plateau in the fidelity decay,
 when the change of $\Delta S_{j} / \hbar $ within the window $W_p$ is negligible.
 The plateau disappears when $\sigma w_p \ww U_{I}' t \sim \pi $.
 If, for example, $U_I=0$ for the system $H_0$,
 then $U_I \propto \epsilon$  for the system $H$,
 and the plateau will end at a time proportional to $\hbar^{1/2} \epsilon^{-2}$
 for $\xi \propto \hbar^{1/2}$ in agreement with the result of \cite{PZ03}.

 Due to the Gaussian term on the right hand side of Eq.~(\ref{mj}),
 the main contribution to $m_{sc}(t)$ comes from $p_{0j}$ not far from $\ww p_0$.
 For these $p_{0j}$, $\nu_j'$ is almost a constant, as well as the value of $S_f$,
 hence, apart from a common phase, $\Delta S_j$ in Eq.~(\ref{mj}) can be approximated by
 $\epsilon U_{Ij}t$.
 For a sufficiently large number of segments, the sum $\sum_j m_j(t)$
 can be replaced by an integral over $p_0$.
 Then, setting $q=p_0-\ww p_0$, we have
 \be m_{sc}(t) \label{mt-int} \simeq \frac{F(t)}{2\pi \sqrt{\pi} w_p} \int_{-\infty}^{\infty} dq
 e^{-q^2/w_p^2} e^{i\sigma t U_I}. \ee
 Expanding $U_I$ to the second order terms in $q$, we have
 \be M_{\rm sc2}(t) \simeq \frac {2c}{\sqrt{4+(w_p^2 \sigma \ww U_I'' t)^2}}
 \exp \left [ \frac { -2  (w_p \sigma \ww U_I' t)^2 }{4+(w_p^2 \sigma \ww U_I'' t)^2 }
  \right ], \label{Mt-sc-2} \ee
 where $U_I'' \equiv \pp U_I' / \pp p_0$ and $c$ a constant with $c \approx 1$ for sufficiently small $\sigma $.

 From equation (\ref{Mt-sc-2}) it is seen that for $(w_p^2 \sigma \ww U_I'' t)^2 \ll 4$,
 the fidelity has a Gaussian decay
 $e^{-( w_p \sigma \ww U_I' t)^2 /2}$,  which agrees with the one given in \cite{PZ02}
 for weak perturbations.
 On the other hand, if $(w_p^2 \sigma \ww U_I'' t)^2 \gg 4$, i.e.
 for larger time $t$,
 $M_{\rm sc2}(t)$ in Eq.~(\ref{Mt-sc-2}) has a power law decay $1/t$.
 In the transition region between the two decays, the fidelity may have an approximate
 exponential decay (see Fig.~\ref{fig-gauss-exp-real}) which can explain the
 exponential-like decay found numerically in \cite{WH05}.

 With further increasing time $t$, higher order terms in the Taylor expansion of $U_I$ will
 become important and will modify the $1/t$ decay.
 In order to evaluate the effect of higher order terms, we divide the interval
 $W_p$ into subintervals labelled by $l$,
 in such a way that inside each of them the linear approximation of $U_I$ can be used.
 Thus their width must be of the order
 $\delta q \simeq 1/\sqrt{|U_I''|\sigma t}$ and their number in
 the region $W_p$ is $l_m \propto \sqrt{\sigma t}$.
 For sufficiently long time $t$, $l_m \gg 1$ and the width of each subinterval is much smaller
 than $W_p$.
 As a result, the Gaussian term in Eq.~(\ref{mt-int}) can be regarded as constant within each subinterval.
 Then, the integral (\ref{mt-int}) reduces to
 $ m_{sc}(t) \simeq \sum_l m_l(t) = \sum_l c_l e^{i\phi_l (t)}/\sigma t$,
 with some coefficients  $c_l$ and phases $\phi_l(t)$. The detailed behavior of this sum depends on
 the properties of the function $U_I$.
 However, one can give estimates in some limiting cases:
 (i) Since
 $  |\sum_l c_l e^{i\phi_l (t)} | \le \sum_l |c_l | \propto (\sigma t)^{1/2}$,
 the slowest decay is $1/\sqrt{\sigma t}$.
 (ii) The fastest decay is obtained when $ |\sum_l c_l e^{i\phi_l (t)} |$ is a constant
 (which may happen due to mutual cancellation of phases $\phi_l$).
 In this case, $m_{sc} \sim 1/\sigma t$.
 (iii) In the case of random phases $\phi_l$, one has
 $|\sum_l c_l e^{i\phi_l (t)} | \propto (\sigma t)^{1/4}$, then, $m_{sc} \sim 1/(\sigma t)^{3/4}$,
 which coincides with the result given in Ref.~\cite{JAB03} for averaged fidelity.
 (For an analysis of Gaussian and power law decay of fidelity,
 which is based on statistics of action difference, see Ref.~\cite{Vanicek04}c).

 Therefore, in general, the decay of  $M(t)$ has the power law dependence
 $(\epsilon t / \hbar )^{-\alpha }$ with $1 \le \alpha \le 2$.
 This is in agreement with numerical results
 in \cite{WH05},
 as well as with our extensive numerical simulations (see, e.g., Fig.4 and the inset of Fig.3).
 The cases of $\alpha \simeq 1$ or 2 have been found quite rare in our simulations.

 It is important to remark that, contrary to chaotic systems for which the decay rate depends on the
 strength but not on the shape of the perturbation, integrable systems lack of such generic
 behavior. This is typical in the general theory of dynamical
 systems and it is due to the peculiarity of integrability.
 In the present case the numerical value of $\alpha $ depends on the particular shape of $U_I$.

 Our approach can also explain an interesting feature of fidelity decay in regular systems
 observed numerically in \cite{WH05}, that is the fact that, in some case, the decay rate may
 decrease with increasing the perturbation strength $\sigma $.
 First we note that Eq.~(\ref{Mt-sc-2}) is valid
 only when $|\Delta S_{j+1} - \Delta S_j|/ \hbar $ is small compared with $\pi $,
 since it is obtained by replacing the sum
 $\sum_j m_j(t)$ by an integral over $p_0$. Since  $U_I'$ is given
 by $(\Delta S_{j+1} - \Delta S_j)/ (p_{0j+1} - p_{0j})\epsilon t$,
 then $|\Delta S_{j+1}-\Delta S_j|/ \hbar  \simeq 2\pi \sigma |U_{Ij}' / \nu_j '|$,
 which may exceed $\pi $ for sufficiently large $\sigma $.
 In this case, we can replace $\Delta S_j / \hbar$ by $ \psi_j \equiv \Delta S_j / \hbar - 2\pi m j $,
 where $m$ is an integer such that $|\psi_{j+1} - \psi_j| \le \pi $.
 Then using the fact that, for $p_{0j}$ close to $\ww p_0$, $ (p_{0j} - \ww p_{0}) \ww \nu ' t \simeq 2\pi j$
 we find that the term $\sigma \ww U_I' $ in Eq.~(\ref{Mt-sc-2}) can be replaced by
 $ \beta \equiv |\sigma \ww U_I' -m \ww \nu '|$.
 With increasing $\sigma $, the value of $\beta $, which gives the decay rate, oscillates
 between 0 and $|\ww \nu ' |/2$,
 with a period $|\ww \nu ' / \ww U_I'|$.
 The results of Fig.~\ref{fig-many-ft} nicely confirm the above
 analysis.

 Finally a word of comment on the comparison of fidelity decay in
 classically regular and chaotic systems.
 As shown in \cite{PZ02}, in sufficiently weak perturbation regime, $\sigma < \sigma_c \propto \sqrt{\hbar }$,
 the Gaussian decay
 $e^{-(w_p \sigma \ww U_I' t)^2/2}$ in the regular case can be faster than
 the fermi-golden-rule decay in the chaotic case.
 This is not surprising as it may look at first sight,
 since in this regime, $w_p \sigma_c \propto \hbar $ can be quite small
 and the fidelity may remain close to 1 for a time comparable to the Heisenberg time $\sim 1/\hbar $, which
 can be quite long.
 Moreover, the above Gaussian decay in the regular case is followed by a power law decay,
 which is slower than the exponential decay in the chaotic case.

 We thank T.~Prosen for useful discussions.

   This work was supported in part by the Academic Research Fund of the National University of Singapore
 and the Temasek Young Investigator Award (B.L.) of DSTA Singapore under Project Agreement No.~POD0410553.
 Support was also given by the EC RTN contract No.~HPRN-CT-2000-0156, the NSA and ARDA under ARO
 contract No.~DAAD19-02-1-0086, the project EDIQIP of the IST-FET programme of
 the EC, the PRIN-2002 ``Fault tolerance,
 control and stability in quantum information processing'',
 and Natural Science Foundation of China Grant No.~10275011.

 \end{document}